\documentclass[aps,prl,twocolumn,superscriptaddress,showpacs,amsmath,amssymb]{revtex4-1}
\usepackage{graphicx,bbm}
\usepackage{bm}
\usepackage{color}
\usepackage{hyperref}

\newcommand{\tomaz}[1]{#1}

\newcommand{\un}[1]{{\underline{#1}}}
\newcommand{\bra}[1]{{\langle #1 \vert}}

\newcommand{\ket}[1]{{\vert #1 \rangle}}

\newcommand{\ZZ}{\mathbb{Z}}

\def\tr{{\,{\rm tr}}}

\def\one{\mathbbm{1}}

\def\one{\mathbbm{1}}

\begin{document}
\title{Identifying Local and Quasilocal Conserved Quantities in Integrable Systems}
\author {Marcin Mierzejewski}
\affiliation{Institute of Physics, University of Silesia, 40-007 Katowice, Poland}
\author{Peter Prelov\v sek}
\affiliation{Faculty of Mathematics and Physics, University of Ljubljana, SI-1000 Ljubljana, Slovenia }
\affiliation{J. Stefan Institute, SI-1000 Ljubljana, Slovenia }
\author{Toma\v z Prosen}
\affiliation{Faculty of Mathematics and Physics, University of Ljubljana, SI-1000 Ljubljana, Slovenia }
\begin{abstract}
We outline a procedure for counting and identifying a complete set of local and 
quasilocal conserved operators in integrable lattice systems.  The method yields a systematic generation of all independent, conserved quasilocal operators related
to time-average of local operators with a support 
on up to $M$ consecutive sites.  As an example we study the anisotropic Heisenberg spin-$1/2$ chain and show that the number of independent conserved operators grows linearly with $M$. Besides the known local operators there exist novel quasilocal conserved quantities in all the parity sectors.
The existence of quasilocal conserved operators is shown also for the isotropic Heisenberg model. 
Implications for the anomalous relaxation of quenched systems are discussed  as well.

\end{abstract}

\maketitle
{\it Introduction.--} 
One expects that the long--time properties of generic systems  are consistent with the Gibbs ensemble \cite{Goldstein2006,Linden2009,Riera2012,polkrev}. It means that the steady states are determined by very few conserved quantities (CQ), e.g., the total energy, the total spin and the particle number.
However, in the integrable models  there exist  a macroscopic number of local CQ  
which can explain some anomalous transport properties \cite{zotos1997}. Various studies  have recently suggested that steady states of integrable systems \cite{Hershfield1993,Natan2006,santos2011,our2013} are fully specified by these local CQ.  
This conjecture is known as the generalized Gibbs ensemble 
\cite{gge,Eckstein2012,Cassidy2011,Gogolin2011,essler2014}  (GGE)   and 
has been well established in systems which can be mapped on noninteracting particles \cite{rigol2014}. 
It is quite evident that an application of this concept or its possible extension to other integrable systems relies on completeness of the set of
CQ  \cite{nongge1,nongge2,nongge3}.

Let us consider the well known class of integrable models of strongly interacting particles on one-dimensional
lattice with $L$ sites. It includes  anisotropic Heisenberg spin-$1/2$ model (XXZ model), the Hubbard model and 
the supersymmetric $t$-$J$ model. In this Letter we focus on the first one while the procedure 
and arguments are generally applicable. For XXZ model there exist a straightforward procedure \cite{tetelman,grabowski}
to construct a set of translationally invariant CQ,  $Q_m= \sum_j q^m_j$,
where $q^m_j$ are local operators spanning at most $m$ sites starting from $j$. Such local CQ commute with the Hamiltonian and with each other.
On the other hand, it has been recognised that these CQ are not enough to account for  the 
dissipationless spin transport \cite{zotos1996,zotos1997, zotos1999,benz,review2007,shastry, herbrych2011, Marko2011,Sirker2009,robin2013,Tomaz2011,my3} neither for absence of decay of other correlations functions \cite{caux,nongge2} nor the kinetic energy \cite{steinigeweg2013}. A partial resolution of the dilemma was in finding novel 
quasilocal conserved quantities (QLCQ), constructed for XXZ model by one of the present authors 
\cite{tomaz_quasilocal11,tomaz_quasilocal13,tomaz_quasilocal14} and others \cite{affleck14},
which have direct overlap  with the spin current operator. These particular QLCQ  have recently been used to 
extend the validity of  GGE  to cases of current-carrying states \cite{nongge1}. 
Furthermore, results on quantum quenches \cite{caux,nongge2}  suggested that new QLCQ (unrelated to $Q_m$) should exist also in other symmetry sectors.

In this Letter we present a systematic procedure for generating local QC and QLCQ.
It is based on the construction of time--averaged operators $\bar A$ emerging from a general local $A$ spanning up to
$M$ sites.  Such operators by construction
commute with $H$ but are in general highly nonlocal and do not have any meaningful thermodynamic limit $L \to \infty$. 
Hence we use an appropriate Hilbert-Schmidt inner product to select the relevant independent QLCQ. For arbitrary model parameters we find clear evidence for
the existence  of QLCQ in the even spin-flip parity sector, while odd parity QLCQ are found only in the easy-plane regime.  
On the other hand, the total number of CQ and QLCQ appears to scale linearly with $M$ being close to the number of CQ in a noninteracting system, $2(M-1)$, 
and far below the maximum number of mutually commuting (diagonal) operators $\propto 2^M$. Such a finding can have important
consequences for relevance of extended GGE

{\it The method.--}  For concreteness we focus on a paradigmatic example of integrable quantum systems, the XXZ model
\begin{equation}
H = J \sum_{j=1}^L \left[ \frac{1}{2} (S_j^+ S_{j+1}^- + S_j^- S_{j+1}^+)  +
\Delta S_j^z S_{j+1}^z \right] , \label{H}
\end{equation}
where $S_j^{\pm,z}$ are spin-$1/2$ operators and $\Delta$ is the anisotropy. 
To avoid  degenerate states we use twisted boundary conditions 
$S^z_{j+L} \equiv S^z_j, S^\pm_{j+L} \equiv e^{\pm i \phi} S^\pm_j$ introducing  flux 
$\phi \ne 0 $ (see Appendix). 



Within the space of all translationally  invariant traceless observables, named as ${\cal A}_L$, we introduce the following 
(Hilbert-Schmidt) inner product
\begin{equation}
(A|B) = \frac{1}{L}\frac{1}{2^L} \tr A^\dagger B, \label{ab}
\end{equation}
which is equivalent to the infinite temperature ($\beta \rightarrow 0$) correlation
$(A|B)=\langle A^\dagger B\rangle_{\beta=0}/L$. The normalisation is chosen such that {\em extensive}   local  observables, like $Q_m$, have finite norm $\| A\|^2 = (A|A)$ in the thermodynamic limit  $L\to\infty$. We consider
finiteness of this norm in the thermodynamic limit  as a general definition of either locality or {\em quasilocality} of an observable $A$ (see also Ref. \cite{olshanii2015}). 
We define a subspace ${\cal A}^m_L$ of local translationally invariant observables with support of size $m$.
In particular, we can define the basis of ${\cal A}^m_L$ 
($m$-local basis) to be composed of operators
\begin{equation}
O_{\un{s}} = \sum_j \sigma^{s_1}_j \sigma^{s_2}_{j+1} \cdots \sigma^{s_m}_{j+m-1}, \label{os}
\end{equation}
where $\sigma_j^z\equiv2S_j^z,\sigma_j^\pm\equiv\sqrt{2}S_j^\pm,\sigma^0_j\equiv\one,\un{s}=(s_1,\ldots,s_m)$,  
$s_j\in\{+,-,z,0\}$ while $s_{1,m}\in\{+,-,z\}$. Note that for given $m$ there are $N_m = 3 \times 4^{m-2} \times 3$ different $O_{\un{s}}$, and 
${\rm dim}\,{\cal A}^m_L=N_m$ for $m\le L/2$.
Definitions (\ref{ab},\ref{os}) imply that operators are 
{\em orthonormal}, i.e.  $(O_{\un{s}}|O_{\un{s'}})= \delta_{\un{s},\un{s}'}$.

Let us define the time-average of operator $A\in {\cal A}^m_L$
\begin{equation}
\bar{A} = \lim_{ \tau \to\infty} \frac{1}{\tau} \int_0^\tau\!\!dt e^{i H t} A e^{-i H t} =
\!\!\!\sum_{n,n'}^{E_n=E_{n'}}\!\!\!\bra{n} A \ket{n'} \ket{n}  \bra{n'}, \label{bara}
\end{equation}
which by construction gives a conserved operator $[H,\bar{A}] = 0$.
In principle, $\bar A$ can be reexpressed, using exact diagonalization $H\ket{n}=E_n\ket{n}$ via Eq.~(\ref{bara}), in terms of operators (\ref{os}). On this basis one could decide whether the operator $\bar A$ is local ($\in {\cal B}^M_L\equiv \bigoplus_{l=1}^{M}{\cal A}^{l}_L$ for some 
$L-$independent $M \ge m$), or quasilocal (with convergent sum of operators $\in {\cal A}^l_L$
with increasing $l$), or generic nonlocal. Such a direct approach is, however, tedious and less transparent and in following we
use a different protocol.

Picking $M > 0$, we calculate the complete set  of $\bar{O}_{\un{s}} $ of 
$D_M = \sum_{m=1}^M N_m = 3\times 4^{M-1}$ (where $N_1=3$) {\em traceless} operators spanning ${\cal B}^M_L$. 
To answer how many of $\bar{O}_{\un{s}}$ are local or might be quasilocal, and are as well independent
we evaluate Hermitian positive-definite $D_M\times D_M$ matrix 
\begin{equation}
K_{\un{s},\un{s}'} = (\bar{O}_{\un{s}}|\bar{O}_{\un{s}'}) =
 (O_{\un{s}}|\bar{O}_{\un{s}'}) 
\label{kmatrix}
\end{equation}
which can be considered as the generalized stiffness matrix  (at $\beta \rightarrow 0$) 
in analogy with the  (spin) current $J_s$ stiffness $D= \beta (\bar{J}_s|\bar{J}_s)$.
Orthonormal eigenvectors $u_{l,\un{s}}$ of matrix $K$ corresponding to eigenvalues $\lambda_l>0$
generate linearly independent conserved operators \tomaz{$Q'_l$}
\begin{equation}
\tomaz{Q'_l}=\sum_{\un{s}} u_{l,\un{s}} \bar{O}_{\un{s}}=\sum_{\un{s}} v_{l,\un{s}} O_{\un{s}}+Q^{\perp}_l
\label{proj1}, 
\end{equation}
where the operator $Q^{\perp}_l$ has support on more than $M$ sites, hence 
$(Q^{\perp}_l|O_{\un{s}})=0$. Calculating the inner product of $\tomaz{Q'_l}$ with $O_{\un{s}'}$
and utilizing Eq. (\ref{kmatrix}) one finds that $v_{l,\un{s}}= \lambda_l u_{l,\un{s}} $.
Substituting this result back into Eq. (\ref{proj1}) and calculating the (squared) norm of operators
on both hand sides one finds that
\begin{equation}
\lambda_l=\lambda^2_l+\| Q^{\perp}_l \|^2.
\end{equation}
Local CQ with support on up to $M$ sites ($ \| Q^{\perp}_l \|=0$) has 
$\lambda_l=1$ strictly independent of $L$.  
Contrary to this, $\lambda_l|_{L\to\infty} > 0$ corresponding to QLCQ ($ \| Q^{\perp}_l \| > 0$) are always smaller than $1$ gradually approaching unity 
with growing $M$. The objective of our study is to establish how the number of CQ and QLCQ depends on $M$.

{\it Symmetries.--} The matrix $K$ can be decomposed in terms of the symmetries of the system. 
Within the XXZ model~(\ref{H}), one CQ is the magnetisation $S^z_{tot}=\sum_j S^z_j$ implying 
that one may consider $S^z_{tot}$ preserving subspace $[S^z_{tot},O_{\un{s}}]=0$, i.e., subset of $O_{\un{s}}$ 
with the constraint that the number  of $s=+$ equals the number of $s=-$ in the sequence $\un{s}$,
reducing the dimensions $D_M$.  More interesting is the spin-flip $\ZZ^2$ symmetry, generated by the 
parity operator $P = \Pi_j (S_j^+ +S_j^-)$. 
$m$-local operator spaces ${\cal A}^m_L$ can then be decomposed into {\em even}($E$)/{\em odd}($O$), 
with the bases generated by sets
$\{O_{\un{s}} \pm P O_{\un{s}} P\}$. Remarkably, {\em all} known local conserved operators $Q_m$ are even, 
while it has been shown recently \cite{tomaz_quasilocal11,tomaz_quasilocal13,tomaz_quasilocal14} 
that odd QLCQ exist for  $|\Delta| < 1$ which determine the properties of the spin current.
Furthermore, XXZ model  is {\em time-reversal} invariant implying that the \tomaz{time-averaging} matrix $K$ is {\em real symmetric} and its 
eigenoperators
$\sum_{\un{s}} v_{l,\un{s}} O_{\un{s}}$
can be classified as {\em real} (R) or {\em imaginary} (I), being spanned separately \tomaz{by} the \tomaz{bases} $O_{\un{s}} + O^\dagger_{\un{s}}$, 
$i (O_{\un{s}} - O^\dagger_{\un{s}})$. Hence one ends up with four orthogonal sectors: $RE, RO, IE, IO$.

{\it Numerical procedure and results.} 
In order to reduce the computational effort we restrict our calculations to the \tomaz{Hilbert} subspace with $S^z_{tot}=0$. 
This requires a straightforward modification of the scalar product in Eq.~(\ref{ab}) where the number of states $2^L$ should be replaced by
${L\choose L/2}$.
Although the set $\{ O_{\un{s}} \}$ is not orthonormal within the chosen subspace, the overlap matrix $N_{\un{s},\un{s}'}=(O_{\un{s}}|O_{\un{s'}})$ remains
real, symmetric and positive--definite. Since it can be diagonalized by an orthogonal matrix $V$
one can introduce an orthonormal set of operators
$ O_\mu=\theta^{-1/2}_{\mu} \sum_{\un{s}} V_{\un{s},\mu} O_{\un{s}}$, where $ \theta_{\mu} $ 
are eigenvalues of $N$. 
The numerical calculations have been performed via full ED of systems with sizes $L=10 - 20$ 
and the boundary condition twisted by $\phi=10^{-4}$.
For given $M$ and $L$ we start the procedure generating first all operators $O_{\un{s}}$ then the orthonormal ones $O_{\mu}$
and  finally the time--averaged  $\bar O_{\mu}$.  At the end the matrix $K_{\mu,\nu}=(\bar{O}_{\mu}|\bar{O}_{\nu})$ 
is evaluated and diagonalized, leading to the eigenvalues $\lambda_l \in [0,1]$. 
The number of non--vanishing eigenvalues is the quantity that we are looking for: the number of local or quasilocal mutually orthogonal conserved operators.

Let us first consider a more generic case as a test of the method. Adding to the XXZ hamiltonian (\ref{H}) a
next nearest neighbor interaction term $H'= \Delta_2~ \sum_j S^z_{j+2} S^z_j$ the model becomes nonintegrable.  
In this case we expect the existence of a single CQ (besides $S^z_{tot}$) which is the full Hamiltonian,
i.e. $Q_2=H+H'$. In Fig.~1 we show the results for eigenvalues $\lambda_l$ vs. $1/L, L=10,\ldots,18$ for 
both $E/O$ sectors, choosing parameters $\Delta=0.5, \Delta_2=0.5$. This
implies the expectation that $\lambda_1=1$ being independent of $L$ for $M\tomaz{\ge}3$, as confirmed
in Fig.~1. All other $\lambda_l$ in both sectors vanish with increasing $L$, predominantly exponentially, 
$\lambda_l \propto \exp(-\zeta L)$. There are exceptions decaying as $\lambda \propto 1/L$ which might
be related to powers of local operators, e.g., $(Q_2)^2$ which are however nonlocal quantities \cite{zotos_unpub}. 

\begin{figure}
\includegraphics[width=0.48\textwidth]{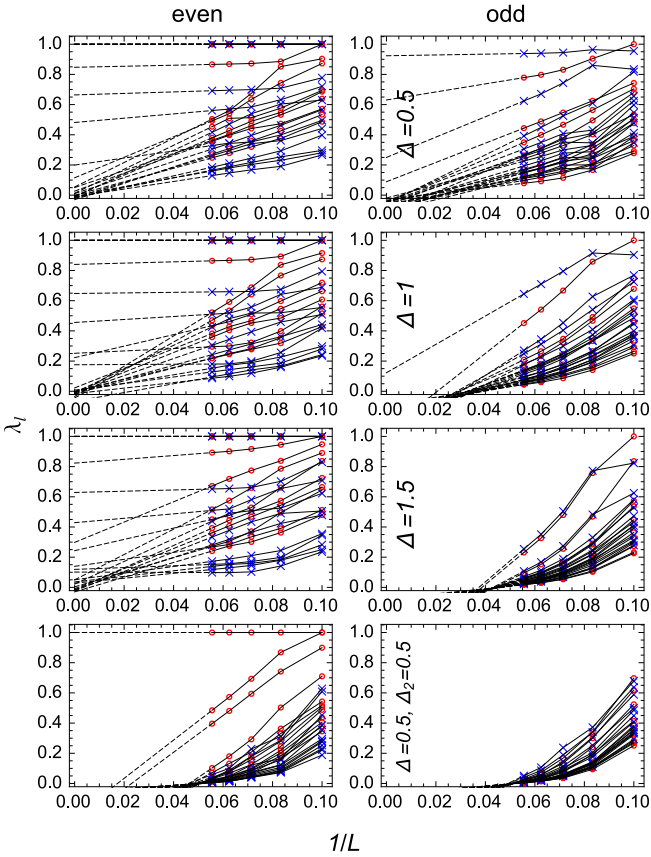}
\caption{(Color online) The size ($1/L$) scaling of leading eigenvalues $\lambda_l$ of matrix $K$ with $M=6$, corresponding to symmetric (red circles) or antisymmetric (blue crosses) eigenoperators with respect to time reversal.
Left/right column shows even/odd parity sectors, while rows indicate different regimes of integrable (upper three rows) and non-integrable (lower row) with parameters indicated in the panel. Dashed lines indicate $1/L$
extrapolation to TL which in some cases provide clear indication of existence of QLCQ $\lambda_l|_{L\to\infty} > 0$, beyond the local eigenoperators with $\lambda_l=1$. }
\label{fig1}
\end{figure}

\begin{figure}
\includegraphics[width=0.48\textwidth]{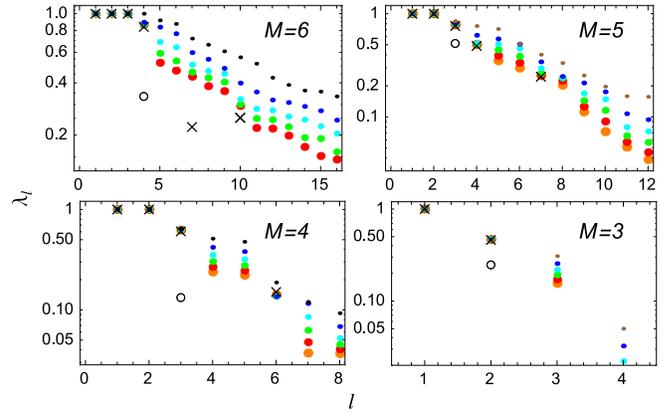} 
\caption{(Color online) 
Dependence of leading eigenvalues $\lambda_l$ of $K$ in $RE$ sector
for isotropic HM $\Delta=1$. Different panels indicate decreasing support sizes $M=6,5,4,3$, while decreasing sizes of points
and colors indicate the system size $L=20$ (orange), $18$ (red), $16$ (green), $14$ (cyan), $12$ (blue), $10$ (brown). Extrapolated $L\to\infty$ values are indicated with crosses if in the range of the plot. Open circles are explained in the text.}
\label{fig2}
\end{figure}

On the other hand, for the integrable XXZ model one finds several CQ and QLCQ as shown in the three topmost rows 
of Fig.\ref{fig1} as well as in  Fig.\ref{fig2}. The latter figure demonstrates in more detail how the spectra of the matrix $K$ depend on $M$ and $L$.
Results in Fig.\ref{fig1} show that the strictly local CQ exist only in the even sector. These are exactly the well known CQ described in Refs. \cite{tetelman,grabowski}.  All the other (novel) CQ are quasilocal. 
In the easy-axis, $\Delta >1$, or isotropic, $\Delta=1$, cases QLCQ exist only in the even parity sector. We have confirmed the latter observation carrying out a finite--size scaling of $\tr K$ for the anisotropy $\Delta \ge 1$   (not shown). We have found a very clear vanishing of $\tr K$ for $L \rightarrow\infty $, in the odd sector at any fixed $M$, what excludes existence of QLCQ in these sectors.  On the contrary, in the easy-plane regime,
$|\Delta| < 1$, QLCQ exist in all parity and time-reversal sectors. 
   
 While the initial operators $ O_{\un{s}} $ as well as the orthonormal ones  $ O_{\mu} $  have support on not more than $M$ sites, the subsequent time--averaging may extend  their support beyond this value. Therefore, before counting the number of QLCQ one should exclude the possibility that $\bar{O}_{\mu}$ can be expanded solely in terms of higher local CQ having support on more than $M$ sites. In order to estimate this contribution we have calculated      
the first nontrivial eigenvalue of $\tilde{K}_{\mu,\nu} = \sum_{m=2}^\infty (O_{\mu}|Q_m)(Q_m|O_{\nu})/(Q_m|Q_m)$  (for computation we truncate at $m=18$), which would agree with $K_{\mu,\nu}$ under the assumption that the orthogonal local operators $Q_m$  are a complete set, i.e. that $\bar{A} = \tilde{A}\equiv \sum_{m} (Q_m|A) Q_m$. These leading
eigenvalues of $\tilde{K}_{\mu,\nu}$, which are shown in Fig. \ref{fig2} as open circles, are always well below eigenvalues of unprojected $K_{\mu,\nu}$ clearly indicating that the set of known $Q_m$ is incomplete also in the even sector.
The central question is how the number of CQ and QLCQ grows with the size of the support $M$ and, in particular, whether this growth is linear 
as in the case of noninteracting particles, where one finds $2M-2$ local operators,
$
Q^R_{m+1} = \sum_k 2\cos(m k) n_k=  \sum_j  (c^{\dagger}_{j+m}c_j+\mathrm{H.c.})$,
$
Q^I_{m+1} = \sum_k 2\sin(m k) n_k= i \sum_j (c^{\dagger}_{j+m}c_j-\mathrm{H.c.})
$,
$m=1,\ldots,M-1$, rewritable into the spin language via
$c^{\dagger}_{j+m}c_l = (1/2)\sigma^+_{j+m}\sigma^z_{j+m-1}\cdots \sigma^z_{j+1} \sigma^-_j$. 

\begin{figure}
\includegraphics[width=0.48\textwidth]{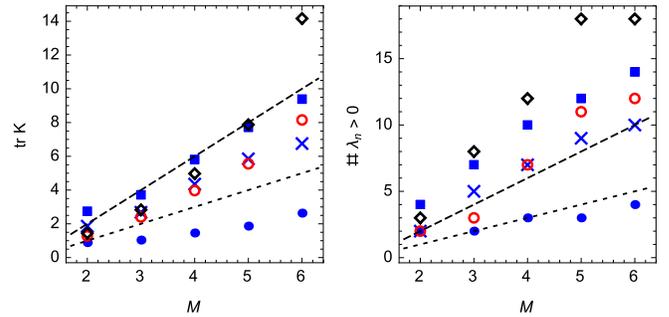}
\caption{
(Color online)
Left: The extrapolated $1/L \to 0$ value of $\tr K$ versus support size $M$ for different regimes of XXZ model: 
$\Delta=1/2$ (blue squares), where contributions for even/odd parity sector are shown separately with crosses/disks, 
$\Delta=1$ (red circles), $\Delta=3/2$ (diamonds).
Right: similar plot for the number of non-vanishing extrapolated eigenvalues $\lambda_l$.
Long/short dashed lines indicate number of known local CQ for interacting $(M-1)$/non-interacting $(2M-2)$ cases.}
\label{fig3}
\end{figure}

 Since the eigenvalues $\lambda_l \le 1$, the total number of CQ and QLCQ 
is obviously bounded from below by tr$K$. Moreover, as all CQ with support on up to $M$ sites correspond to $\lambda_l=1$, $\tr K$ should gradually approach the number
of QC and QLCQ for large enough $M$. In Fig. \ref{fig3} we show $\tr K$ obtained after the $1/L$ size scaling. For comparison we show also
the number of nonzero $\lambda_l$ counted directly after carrying out a finite--size scaling of individual leading eigenvalues. 
The number of directly counted CQ and QLCQ as well as $\tr K$ increase linearly with $M$. 
We find it particularly surprising and suggestive that the in easy--plane regime $\tr K$ is very close to $2(M-1)$, i.e. to the number of CQ in the systems of
noninteracting particles. However, contrary to the latter systems tr$K$ in XXZ model is not equaly distributed among the parity sectors. 
Half of the total tr$K$ comes from the known (even) QC, approximately one quarter originates from even QLQC and one quarter from odd QLCQ.
In the isotropic and Ising regimes ($\Delta \ge 1 $) QLQC in the odd sector disappear, while the total number of CQ and QLCQ in  the even sector is evidently 
larger than $(M-1)$ in noninteracting case. Our central observation concerns the main difference between interacting and noninteracting 
integrable systems. We have found that this difference does not consists in the number of QC (which is extensive in both cases) but rather in their locality and symmetry.


{\it Identifying the conserved quantities.--} 
Besides known $Q_m, m=2,3\ldots$ which are all in the even sector
we did not find any other strictly local CQ. As expected we also confirmed the existence
of the QLCQ in the odd sector which has been analytically constructed 
\cite{tomaz_quasilocal11,tomaz_quasilocal13,tomaz_quasilocal14,affleck14} and
used in the extended GGE previously \cite{nongge1}.
Furthermore, very weak double scaling of eigenvectors of the matrix $K$ with, first, increasing $L$, and then, increasing $M$, allows us to clearly identify novel QLCQ.
In the isotropic case $\Delta=1$ and ER symmetry sector, the first nontrivial QLCQ $Q'$ reads approximately, writing all terms to order $m=4$,
$Q' \propto \sum_j ({\bf S}_{j+2}\cdot {\bf S}_j + {\bf S}_{j+3}\cdot{\bf S}_{j} + ({\bf S}_{j+3}\cdot {\bf S}_{j+2}) ({\bf S}_{j+1}\cdot {\bf S}_j) - 2 ({\bf S}_{j+3}\cdot {\bf S}_{j+1}) ({\bf S}_{j+2} \cdot{\bf S}_j)  )+...$.
See Appendix for details.


{\it Steady state  after a linear quench.--} Let us assume that at time $t=0$ the Hamiltonian is quenched  from 
$H-X_a$ to the integrable  $H$, 
where $X_a\in {\cal A}_L$ is considered as a perturbation. The system is initially in the Gibbs state $\rho_a=\exp[-\beta(H-X_a)]/Z$, 
where $Z= \mathrm{Tr} \exp[-\beta(H-X_a)] $, and it relaxes towards a steady state  
$\overline{\rho}_a$.
Since $\overline{X}_a$ commutes with $H$ one easily finds the following linear expansion  
\begin{equation}
\overline{ \exp[-\beta(H-X_a)]}  \simeq    \exp(-\beta H)(1+\beta \overline{X_a}).
\label{bar}
\end{equation} 
Calculating the trace over the eigenstates of $H$ 
one obtains an analogous approximation for the partition function 
$Z= Z_T (1+\beta \langle X_a \rangle )$.
Here $Z_T= \mathrm{Tr}[ \exp(-\beta H)]$  and the average $\langle ... \rangle $ is defined for the thermal state 
$ \rho_T= \exp(-\beta H)/Z_T$. At the end we obtain the linear approximation for the steady state:
\begin{equation}
\overline{\rho}_a= \rho_T[1+\beta(\tomaz{\overline{X}}_a- \langle X_a \rangle )]
\end{equation}
Next, we consider some other extensive observable $X_b\in{\cal A}_L$ and a related intensive one $X_b/L$. We study whether the (possibly) nonthermal $\overline{\rho}_a$  and the thermal $\rho_T$ states can be distinguished 
by the measurement of $X_b/L$. We calculate 
\begin{equation}
\beta K_{ab}  = \mathrm{Tr}\left[(\overline{\rho}_a-\rho_T)\frac{X_b}{L}\right] =
  \frac{\beta}{L}(\langle \tomaz{\overline{X}_a \overline{X}_b} \rangle - \langle  \overline{X}_a \rangle  \langle \overline{X}_b \rangle ).          
\end{equation} 
In the high--temperature regime ($\beta \rightarrow 0$) and for traceless operators $X$
the above correlation matrix $K$ coincides with the matrix of inner products in Eq. (\ref{kmatrix}).     

{\it Conclusions.--}
We have presented a systematic procedure which allows to establish the existence of local and
quasi local CQ in 1D many-body models with short range interactions.  In spite of limitations of our results
to finite sizes $L\leq 20$, explicitly performed only within
the (unpolarised) subspace $S^z_{tot}=0$,  they allow for some firm conclusions for XXZ model. The method
confirms besides the known strictly local CQ, being all in the even sector, also even and odd
 QLCQ. In 
the metallic regime $\Delta<1$ the odd QLCQ are consistent with the finite
spin stiffness $D > 0$ and analytical construction  
\cite{tomaz_quasilocal11,tomaz_quasilocal13,tomaz_quasilocal14,affleck14}. 
The novel QLCQ in the even sector exist in the whole $\Delta>0$ range and 
can explain at least part of deviations from GGE observed so far in quenched spin systems at $\Delta > 1$  \cite{caux,nongge2}.
There are clear  indications for the existence of further QLCQ which emerge with increasing support size $M$ but for $M\leq6$ are not yet suffiently converged  to determine their explicit form. 
It is nevertheless plausible  that in quenched or driven systems the major role in thermalisation will be related to CQ and QLCQ with smaller 
supports $M$ as identified in our study.  
Here the most important conclusion is that the number of local and quasilocal CQ appears to scale linearly with $M$, for $|\Delta| < 1$, being approximately
$2(M-1)$ similar to a model of noninteracting fermions.  
Our results allow for a meaningful extension of GGE  incorporating the full set of QLCQ.

\acknowledgements
M.M. acknowledges support from the  DEC-2013/09/B/ST3/01659 project of the Polish National Science Center. 
P.P. and T.P. acknowledge support by the program P1-0044 and projects J1-4244 (P.P.) and J1-5349, N1-0025 (T.P.) 
of the Slovenian Research Agency.

\section*{APPENDIX}

\begin{figure}
\includegraphics[width=0.45\textwidth]{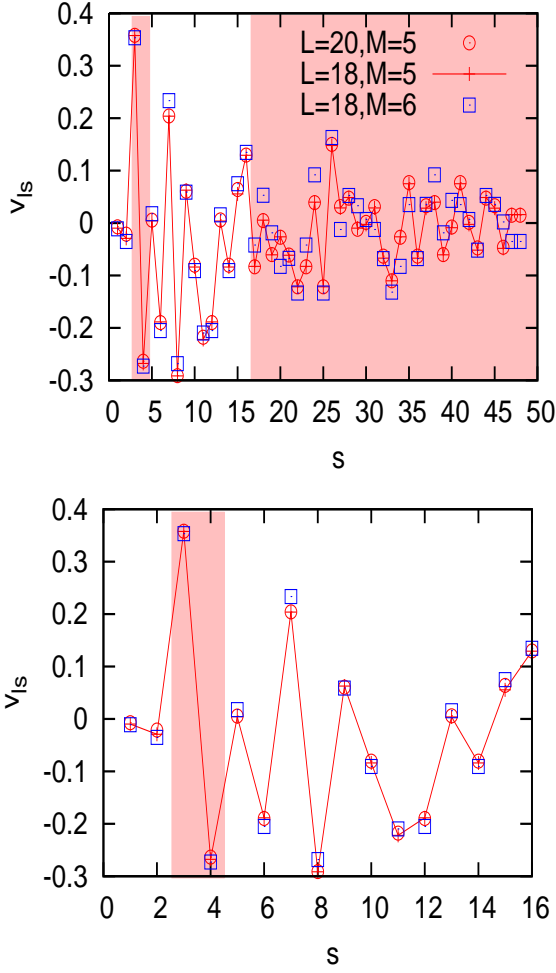}
\caption{Projections of $Q'$ ($\Delta=1$,  RE sector) on the basis vectors $O_s$ as defined  in Eq. (6) in the main text. The explicit form of $O_s$ for $s=1,..,16$ is shown in the text. The shaded/unshaded backgrounds from the left to right side of figure mark the sets of $O_s$ with support on $2,3,4$ and 5 sites, respectively. Results for $s \le 16$ are
magnified in the lower panel.}
\label{fig4}
\end{figure}

In the Appendix we give some technical details of our study of the leading local and quasilocal conserved quantities (QLCQ) of the isotropic Heisenberg model. We also  demonstrate that while the twisted boundary conditions facilitate the numerical calculations by lifting degeneracy of eigenstates, they have negligible influence on the numerical results, provided that the related flux $\phi$ is small enough.

{\em Leading QLCQ.}--
Here we study the leading QLCQ in the real even--parity ($RE$) sector  of the isotropic Heisenberg model ($\Delta=1$) and we denote this QLCQ by $Q'$.  The norm of $Q'$ is shown in Fig. 2 in the main text as the 4th eigenvalue for $M=6$ as well as 
the 3rd eigenvalue for $M=5$ or $M=4$.   In Fig. \ref{fig4} we show projections, $v_{ls}$, of $Q'$ on the basis operators $O_s$ as defined in Eq. (6) in the main text. For the sake of completeness we list also explicit form of all $O_s$ with support on up to 4 sites:    
 \begin{eqnarray}
O_1&=& \sqrt{2} \sum_i (  S^-_i S^+_{i+1}+H.c. )  \\
O_2&=& 4 \sum_i S^z_iS^z_{i+1}   \\
O_3&=& - \sqrt{2} \sum_i ( S^-_i S^+_{i+2}+H.c.) \\
O_4&=& 4 \sum_i S^z_iS^z_{i+2}   \\
O_5&=&  4 \sqrt{2} \sum_i (  S^-_i S^z_{i+1} S^z_{i+2} S^+_{i+3}+H.c. )  \\
O_6&=& -4 \sqrt{2} \sum_i ( S^-_i S^z_{i+1} S^+_{i+2} S^z_{i+3}+ H.c. )   \\
O_7&=&  2 \sqrt{2} \sum_i ( S^-_i S^+_{i+2} S^-_{i+1}  S^+_{i+3}+ H.c.)    \\ 
O_8&=& \sqrt{2}\sum_i S^-_i S^+_{i+3}+H.c.   \\
O_9&=& -2 \sqrt{2} \sum_i( S^-_i S^+_{i+1} S^-_{i+2}  S^+_{i+3}+ H.c.)    \\
O_{10}&=& 4\sqrt{2} \sum_i  S^-_i S^+_{i+1} S^z_{i+2}  S^z_{i+3}   \\
O_{11}&=& 4 \sum_i  S^z_i S^z_{i+3} +H.c.   \\
O_{12}&=& - 4 \sqrt{2}  \sum_i( S^z_i S^-_{i+1} S^z_{i+2}  S^+_{i+3}+H.c. )   \\
O_{13}&=& 4 \sqrt{2} \sum_i ( S^z_i S^-_{i+1} S^+_{i+2}  S^z_{i+3}+H.c. )  \\
O_{14}&=& 4 \sqrt{2} \sum_i (S^z_i S^z_{i+1} S^-_{i+2}  S^+_{i+3}+H.c. )  \\
O_{15}&=& 16 \sum_i  S^z_i S^z_{i+1} S^z_{i+2}  S^z_{i+3}   \\
O_{16}&=& 2\sqrt{2}  \sum_i ( S^+_i S^-_{i+1}  S^-_{i+2}  S^+_{i+3} +H.c )    
 \end{eqnarray}
 
 In order to establish $Q'$ in the thermodynamic
 limit one should carry out two subsequent finite--size scalings of $v_{ls}$: first $L \rightarrow \infty$ and then  $M \rightarrow \infty$. Results shown in the upper panel in Fig. \ref{fig4} indicate that the range of accessible $L$ and $M$ is insufficient to carry out such scaling for all $s$.  However, the projections of $Q'$ on the basis operators with support on up to $4$ sites ($s=1,...,16$) only weakly depend on $M$ and $L$, as shown in the lower panel in the same figure. We stress that the most important (mostly studied) physical quantities have support on a few lattice sites only. In order to study the influence of $Q'$ on the dynamics of these quantities in the high--temperature regime, it is sufficient to specify $v_{ls}$ only for these $Q_s$ which have the same support as the investigated quantities. Hence, the projections shown in the lower panel 
in Fig. \ref{fig4} are of key importance.

In order to get a deeper insight into the structure of QLCL we are focusing on the first nontrivial $Q'$ in the case of SU(2) symmetric (isotropic) $XXX$ model ($\Delta=1$). Using a simple group-theoretic argument one realises that the operators $Q'$ should also be SU(2) invariants. In fact, they should be SU(2) scalars, unless they correspond do degenerate eigenvalues of the matrix $K$ -- which we have checked that it never takes place for finite $M$.
As the only translationally invariant SU(2) scalar is the Hamiltonian itself $H = J \sum_j \bf{S}_{j+1}\cdot\bf{S}_j$, this immediately implies that the projections
of $Q'$ on $O_1$ and $O_2$ should be vanishingly small, in the limit $L\to\infty$, as clearly demonstrated in Fig. \ref{fig4}. 
We note that $SU(2)$ invariance is expected to be restored only in the thermodynamic limit  as we work with the microcanonical ensemble of zero total spin projection $\sum_j S^z_j$, which is (strictly) not SU(2) invariant for finite $L$.
In other words, nonvanishing projections on $O_1$ or $O_2$ would imply that ether $Q'$  is not SU(2) invariant or it is not orthogonal to the Hamiltonian $\sum_i {\bf S}_{i+1} {\bf S}_i $. The general form of SU(2)--invariant $Q'$  with all terms  up to order $m=4$
\begin{eqnarray}
Q' &=&  \sum_i \left[  \alpha {\bf S}_{i+2}\cdot{\bf S}_i +  \beta ({\bf S}_{i+3}\cdot{\bf S}_{i+2}) ({\bf S}_{i+1}\cdot{\bf S}_i) \right. \nonumber \\ 
&&+ \gamma ({\bf S}_{i+3}\cdot{\bf S}_{i+1}) ({\bf S}_{i+2}\cdot{\bf S}_i) + \nonumber \\
&& \left. +\delta ({\bf S}_{i+3}\cdot{\bf S}_{i}) ({\bf S}_{i+2} \cdot {\bf S}_{i+1}) + \zeta ({\bf S}_{i+3}\cdot{\bf S}_{i}) \right]+..., \nonumber \\
\label{eqs1}
 \end{eqnarray}
 can be expressed in terms of basis operators $O_s$
 \begin{eqnarray}
Q' &= &  \frac{1}{4}(\alpha O_4-\sqrt{2} \alpha O_3 + \zeta O_{11} +\sqrt{2} \zeta O_{8}) \nonumber \\   
  &  &+  \frac{\sqrt{2}}{16}[ (\delta+\gamma)O_7-(\beta+\delta) O_9+(\beta+\gamma)O_{16}] \nonumber \\     
&&+\frac{\sqrt{2}}{16}[\beta(O_{10}+O_{14})-\gamma(O_{6}+O_{12}) \nonumber \\
&& + \delta(O_{5}+O_{13})]  +\frac{1}{16}(\beta+\delta+\gamma)O_{15}.
\label{eqs2}  
\end{eqnarray}
Numerical results for $v_{l3},...,v_{l16}$ shown in Fig. \ref{fig4} 
show that Eq. (\ref{eqs2}) indeed holds true confirming SU(2) invariant
form of $Q'$ for isotropic Heisenberg model.   
We have also found the following approximate values of parameters in Eq. (\ref{eqs1}):
\begin{eqnarray}
\alpha  & \simeq & \beta \simeq \zeta, \\
\gamma & \simeq & -2 \alpha, \\
|\delta| & \ll & |\alpha|. 	 
\end{eqnarray}

\begin{figure}
\includegraphics[width=0.45\textwidth]{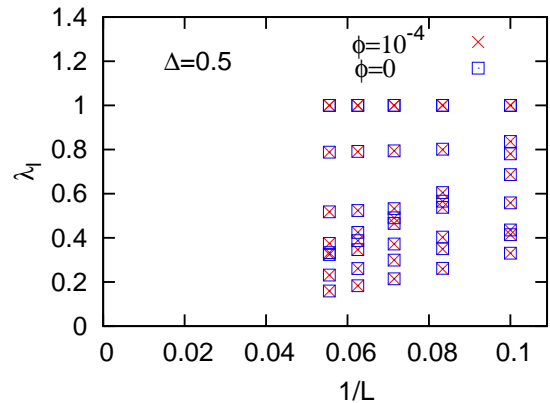}
\caption{ The size ($1/L$) scaling of leading eigenvalues $\lambda_l$ of matrix $K$ with $M=5$ obtained for RE sector of anisotropic Heisenberg. Fig. 1 in the main text shows analogous results for $M=6$. Crosses and squares show results obtained for twisted and period boundary conditions, respectively. } 
\label{fig5}
\end{figure}    

{\em Twisted boundary conditions}.--
In order to avoid degeneracy of energy levels we used in the computations twisted boundary conditions $S^z_{j+L} \equiv S^z_j, S^\pm_{j+L} \equiv e^{\pm i \phi} S^\pm_j$ introducing  flux  $\phi = 10^{-4} $. As immediately follows from Eq. (4) in the main text, $\phi\ne 0$  facilitates the numerical calculations focused on the time-averaged operators. In Fig. (\ref{fig5}) we compared results obtained for $\phi=0$ and $\phi=10^{-4}$ which confirm that such small values of $\phi$ 
have negligible influence on the eigenvalues of the $K$ matrix. 

\bibliography{bib_nongge.bib} 

\end{document}